\documentclass[ amsmath, amssymb,10pt,sort&compress ]{article}
\usepackage{amsmath,amssymb,graphicx}
\usepackage{xcolor}
\usepackage{hyperref}
\usepackage{cleveref}
\usepackage{filecontents}
\usepackage{authblk}
\usepackage{graphicx}
\usepackage{caption}
\usepackage{subcaption}
\usepackage{float}
\usepackage{xcolor}
\usepackage[utf8]{inputenc}
\usepackage{blindtext}
\usepackage{setspace}
\restylefloat{figure}
\input{epsf.sty}
\newcommand{\nc}{\newcommand}
\nc{\ba}{\begin{eqnarray}}
\nc{\ea}{\end{eqnarray}}
\newcommand\be{\begin{equation}}
\newcommand\ee{\end{equation}}

\title{Tachyon Inflation  in $R+R^2$ Background}
\author[1]{Salomeh Khoeini-Moghaddam\thanks{skhoeini@khu.ac.ir}}
\affil[1]{\tiny{Department of Astronomy and High Energy Physics, Faculty of Physics, Kharazmi University, Mofateh Ave., Tehran 15719-14911, Iran}}
\date{}
\begin{document}
\maketitle
\begin{abstract}
\small{The presence of a tachyon field in $R+\mu R^2$ is considered. Our analysis shows that the contribution of the tachyon field in energy density is suppressed, but it affects cosmological parameters.}
%{\bf Keywords:Inflation; Tachyon; Starobinsky }
\end{abstract}
%%%%%%%%%%%%%%%%%%%%%%%%%%%%%%%%%%%%%%%%%%%%%%%%%%%%%%%%%%%%%%%%%%%%%%%%%%%%%%%%%%%%%
\section{Introduction}
Inflation theory \cite{Guth:1980zm,Linde1,Albrecht}is the dominant paradigm for the early universe, which generates super-horizon fluctuations with an almost scale-invariant spectrum. This theory is in good agreement with observations\cite{Planck2018b}. There are a large number of models of inflation. The slow-roll model is the simplest, in which a scalar field rolls down in an almost flat potential.
There are many ways to modify this simple model. One way is to consider fields with nontrivial kinetic terms. These kinds of models are called k-inflation\cite{mukhanov1}.

String theory inspires some strange fields. One of these fields is the tachyon, having a nontrivial kinetic term. Tachyon fields are the lowest energy state in the unstable brane-untibrane and Dp-brane systems\cite{sen}. These fields can produce rapid expansion, so they are considered a candidate for dark energy and inflation.
On the other hand, in string theory,  $\alpha'$ correction permits the existence of a higher order derivative of the gravitational field \cite{Wang2016} (and references therein). This issue motivates the proposition of modified gravitational models. The $f(R)$ model is a simple extension of Einstein's gravity, and the Starobinsky model is the most famous one\cite{Staro1}\cite{Staro2}. First,this model proposed to derive inflation without any extra field. In $f(R)$ models, by transforming from the Jordan frame to the Einstein frame, we get an additional scalar field called scaleron\cite{witt}.

Although, in the original Starobinsky model, the scalaron causes inflation by itself, there is no logical reason that other fields should not be considered; the presence of other fields rather than the scalaron is assumed in many works \cite{Bamba:2015uxa,Myrzakulov2015,Canko_2020,Gomes_2017,Mori17, Elizalde_2019,Ketov2020-3, Antoniadis19,Ketov2020-1,Bruck2010,Bruck2011,khoeini20}.
 With additional fields in the $f(R)$ background, we arrive at multi-field inflation in the Einstein frame. In these models, fields can interact with each other, which affects the inflation dynamics and also can cause some features.We are interested in studying tachyon fields in the context of the Starobinsky model. The existence of the tachyon field in f(R) theory is investigated in several works\cite{Wang2016, Jamil2011}.The interaction of tachyon in Einstein's gravity is studied in previous works \cite{interaction1, interaction2, interaction3}. In this work, we consider a tachyon field in the $R+\mu R^2$ background and do a study on inflation in the early universe.

The structure of this paper is as follows: in section (\ref{sec-model}), we present our model; in section (\ref{sec-back}), we analyze the evolution of the fields in FRW metric. We investigate the perturbation of fields and observational parameters in section(\ref{sec-perturb}). Finally, we conclude in section(\ref{sec-conclude}).
%~~~~~~~~~~~~~~~~~~~~~~~~~~~~~~~~~~~~~~~~~~~~~~~~~~~~~~~~~~~
\section{The Model}\label{sec-model}
The action of a tachyon field in general $F\left(R\right)$ background can be written as
\begin{align}\label{action1}
S &=\frac{1}{2\kappa^2}\int d^4x\sqrt{-g}f\left(R\right)-\int d^4x\sqrt{-g}V\left(\phi\right)\sqrt{1-2\alpha'X},
\end{align}
where $X = -\frac{1}{2}g_{\mu\nu}\partial^\mu\phi\partial^\nu\phi$, is kinetic term and $\alpha'$ is coupling constant, we set $\alpha'=1$. $\kappa^2=8\pi G=M_{pl}^{-2}$($c=1$); in natural unit we assume $\kappa^2=1$. We rewrite the action as below
\begin{align}\label{action2}
  S &=\int d^4x\sqrt{-g}\left(\frac{1}{2\kappa^2}f'R-\frac{1}{2\kappa^2}f'R-\frac{1}{2\kappa^2}f\right)-\int d^4x\sqrt{-g}V\left(\phi\right)\sqrt{1-\alpha'X},
\end{align}
the last two terms in the first integral can be interpreted as a potential term  $U =\frac{f'R-f}{2\kappa^2}.$
It is viable to transform to the Einstein frame by conformal transformation,$\tilde{g}_{\mu\nu} =\Omega^2=e^{2\beta\psi}g_{\mu\nu}.$
We adopt the conformal factor as $\Omega^2=e^{2\beta\psi}=f'$, which allows us to write the action in this frame as below,
\begin{align}\label{action3}
  S'&= \int d^4x\sqrt{-\tilde{g}}\left(\frac{1}{2\kappa^2}\tilde{R}-\frac{6\beta^2}{2\kappa^2}\tilde{g}^{\rho\sigma}\tilde{\nabla}_\rho\psi\tilde{\nabla}_\sigma\psi-w\left(\psi\right)\right) \\\nonumber
    &-\int d^4x\sqrt{-\tilde{g}}e^{-4\beta\psi} V\left(\phi\right)\sqrt{1+\alpha'e^{2\beta\psi\tilde{g}^{\rho\sigma}\tilde{\nabla}_\rho\phi\tilde{\nabla}_\sigma\phi}},
\end{align}
where  $\tilde{}$ refers that the quantities are written down with respect to the new metric, the constant $\beta$ is defined as $\beta=\sqrt{\frac{1}{6}}\kappa$.
The potential  term can be written as a function of $\psi$; its exact form depends on the form of $f\left(R\right)$. We choose the Starobinsky gravity, $f\left(R\right)=R+\mu R^2$, that the potential has a simple form, $w\left(\psi\right)=\frac{1}{8\kappa^2\mu}e^{-4\beta\psi}\left(e^{2\beta\psi}-1\right)^2$(we denote the potential with "w").
In the rest of the paper, we assume the Starobinsky model coupling satisfies the condition, $\mu\gg \kappa^2$. Based on  CMB amplitude observations, we set $\mu\sim10^9M_{pl}^{-2}$.
%%%%%%%%%%%%%%%%%%%%%%%%%%%%%%%%%%%%%%%%%%%%%%%%%%%%%%%%%%%%%%%%%%%%%%%%%%%%%%%%%%%%%%%%%%%%%%%%%%%%%%%%%%%%%%%%%%%%%%%%%%%%%%%%%%%%%%%%%%%%%%%%%%%%%%%%%%%%%%%%%%%%%%%%%%%%%%%%%%%%%%%%%%%%%%%%%%%%%%%%
\section{The Background Evolution}\label{sec-back}
 We assume that the metric in  the Einstein frame is flat FRW, $d\tilde{S} = -d\tilde{t}^2+a^2\left(t\right)d\vec{\tilde{x}}^2$.
 In this background the fields are only functions of t i.e. $\psi=\psi\left(t\right)$ and $\phi=\phi\left(t\right)$. The equations of motions for $\psi$ and $\phi$ are
\begin{align}
  \ddot{\psi}&+3\frac{\dot{a}}{a}\dot{\psi}+w_{,\psi}-A\beta e^{-4\beta\psi}V\left(\phi\right)\left(1+3A^{-2}\right)=0 \label{eqm1psi}  \\
  \ddot{\phi}&+ 3A^{-2}\frac{\dot{a}}{a}\dot{\phi}+\beta\left(1-3A^{-2}\right)\dot{\psi}\dot{\phi}+\frac{A^{-2}}{1-A^{-2}}\frac{V_{,\phi}}{V}\dot{\phi}^2=0,\label{eqm1phi}
\end{align}
where we have defined $A=\frac{1}{\sqrt{1-\alpha'e^{2\beta\psi}\dot{\phi}^2}}$.

The Tachyon potential must satisfy the following conditions\cite{rezazadeh2017, steer2004}, $V\left(\phi\rightarrow\infty\right)\rightarrow0$ and $\frac{dV}{d\phi}<0$,
we choose the potential form  as below
 \begin{align}\label{phi-potential}
   V\left(\phi\right)=&V_0\left(1+\frac{1}{27}\left(\frac{\phi}{\phi_0}\right)^4\right)e^{-\left(\frac{\phi}{\phi_0}\right)}
 \end{align}
where $V_0$ and $\phi_0$ are constants\cite{rezazadeh2017,steer2004,Vulcanov}. For Tachyon models in Einstein gravity, this potential is compatible with Planck data\cite{rezazadeh2017}.
We can write the energy density and pressure for $\psi$ and $\phi$ as below:
\begin{align}\label{rho-p}
  \rho_{\psi}&=\frac{1}{2}\dot{\psi}^2+w\left(\psi\right) \hspace{1cm} p_{\psi}=\frac{1}{2}\dot{\psi}^2-w\left(\psi\right) \\
  \rho_{\phi} &=e^{-4\beta\psi}V\left(\phi\right)A \hspace{1cm} p_{\phi}=-e^{-4\beta\psi}V\left(\phi\right)A^{-1}.
\end{align}

The total energy density and pressure are the sum of individual quantities. The numerical analysis shows that the contribution of the tachyon in the total energy density is suppressed by the scalaron. Therefore, the scalaron governs the dynamics of the Hubble constant and the expansion rate of the universe. This suppression occurs by a factor of $e^{-4\beta\psi}$.

%%%%%%%%%%%%%%%%%%%%%%%%%%%%%%%%%%%%%%%%%%%%%%%%%%%%%%%%%%%%%%%%%%%%%%%%%%%%%%%%%%%%%%%%%%%%%%%%%%%%%%%%%%%%%%%%%%%%%%%%%%%%%%%%%%%%%%%%%%%%%%%%%%%%%%%%%%%%%%%%%%%%%%%%%%%%%%%%%%%%%%%%%%%%%%%%%%%%%%%%%
\section{Perturbation}\label{sec-perturb}
\begin{figure*}[t!]
 \begin{subfigure}[t]{0.6\textwidth}
  \includegraphics[height=3.2cm]{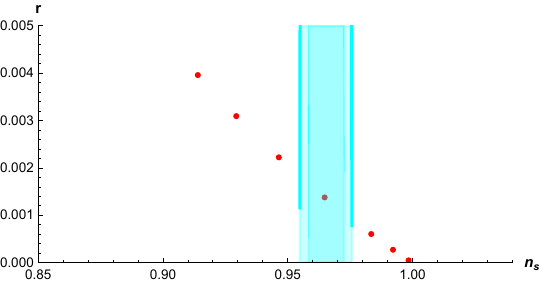}
 \end{subfigure}
 \begin{subfigure}[t]{0.6\textwidth}
    \includegraphics[height=3.2cm]{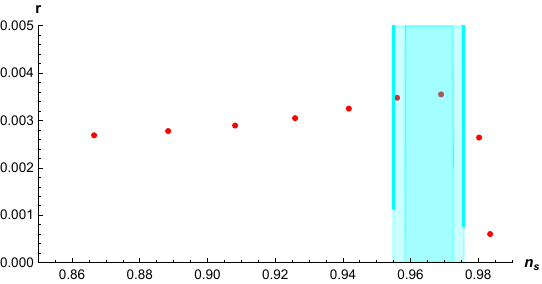}
 \end{subfigure}
\caption{\small{We show  $n_s$-$r$ plane, the color area is Plank2018 data; the red dots are the numerical values for different parameters $v_0$ and $\phi_0$.In the left panel, we set $\phi_0=100$ and vary $v_0$ in the range 1 to 100, and in the right panel, we consider  $v_0=10$ and let $\phi_0$ go from 50 to 500.}}
\label{Fig5}
\end{figure*}

We consider the linear perturbation. We perturb the action (\ref{action3}) by separating the fields into two parts, a homogeneous part which evolves by (\ref{eqm1phi}) and (\ref{eqm1psi}), and a perturbed part, $\psi\left(t,\mathbf{x}\right)= \psi\left(t\right)+ \delta\psi\left(t,\mathbf{x}\right)$ and $\phi\left(t,\mathbf{x}\right)= \phi\left(t\right)+ \delta\phi\left(t,\mathbf{x}\right)$.
We work in the longitudinal gauge; in the absence of anisotropic  stress, the perturbed metric can be written as $ds^2=-\left(1+2\Phi\right)dt^2+a^2\left(t\right)\left(1-2\Phi\right)\delta_{ij}dx^idx^j$.
By varying the action with respect to the perturbed fields, we get the perturbed field equations. We write these equations in Fourier space,
\begin{eqnarray}
  \delta\ddot{\psi}&+& 3H\delta\dot{\psi}-4\dot{\psi}\dot{\Phi}+\delta\psi[\frac{k^2}{a^2}+w,_{\psi\psi}+7\beta^2e^{-4\beta\psi}V\left(\phi\right)\left(A^{-1}+A\right)]\\\nonumber
  &+&2\Phi[w_{,\psi}-\frac{1}{2}\beta e^{-4\beta\psi}V\left(\phi\right)\left(A^{-1}+3A\right)]+\delta\phi[\beta e^{-4\beta\psi}V\left(\phi\right)\left(3A^{-1}+A\right)]\\\nonumber
  &+&\frac{\delta\dot{\phi}}{\dot\phi}[\beta e^{-4\beta\psi}V\left(\phi\right)\left(3A-5A^{-1}\right)]=0 \\\label{perturbed-eqm1}
  \delta\ddot{\phi}&+&[A^{-2}\frac{k^2}{a^2}-A^{-2}\left(1-A^{-2}\right)^{-1}\left(\frac{V^2_{,\phi}}{V^2}-\frac{V_{,\phi\phi}}{V}\right)\dot{\phi}^2]\delta\phi-[2\frac{V_{,\phi}}{V}\dot{\phi}^2\\\nonumber
  &-&7\left(1-A^{-2}\right)\beta\dot{\psi}\dot{\phi}+3\left(2-3A^{-2}\right)H\dot{\phi}]\delta\dot{\phi}+A^{-2}\left(A-3\right)\beta\dot{\phi}\delta\dot{\psi}\\\nonumber
  &-&A^{-2}\left(A^2+3\right)\dot{\phi}\dot{\Phi}-\left(1-A^{-2}\right)^{-1}[2\left(1-A^{-2}+3A^{-4}\right)\frac{V_{,\phi}}{V}\dot{\phi}^2\\\nonumber
  &-&6\left(1-A^{-2}\right)^2\beta\dot{\phi}\dot{\psi}+6\left(1-A^{-2}\right)^2H\dot{\phi}]\beta\delta\psi-\left(1-A^{-2}\right)^{-1}[-2\frac{V_{,\phi}}{V}\dot{\phi}^2\\\nonumber
  &+&3\left(1-A^{-2}\right)^2\beta\dot{\phi}\dot{\psi}-6\left(1-A^{-2}\right)^2H\dot{\phi}]\Phi=0\label{perturbed-eqm2}
\end{eqnarray}
The perturbed Einstein equations give, $\dot{\Phi}+H\Phi =\frac{1}{2}\left(\dot{\psi}\delta\psi+\left(1-A^{-2}\right)Ae^{-4\beta\psi}V\frac{\delta\phi}{\dot{\phi}}\right)$.
We the gauge-invariant Sasaki-Mokhanuv variables, $Q_\psi\equiv\delta\psi+\frac{\dot{\psi}}{H}\Phi$ and $Q_\phi\equiv\delta\phi+\frac{\dot{\phi}}{H}\Phi$,
the co-moving curvature perturbation and its power spectrum can be written as
 \begin{align}\label{curnature-perturb}
  \mathcal{R} & =\frac{H}{-2\dot{H}}[\dot{\psi}Q_\psi+e^{-2\beta\psi}A\dot{\phi}Q_\phi],\\
  \mathcal{P}_\mathcal{R}&=\frac{k^3}{2\pi^2}|\mathcal{R}|^2.
 \end{align}
 We also define two auxiliary fields, $u_\psi\equiv a Q_\psi$ and $u_\phi\equiv ae^{-\beta\psi}A^{2/3}Q_\phi$.
One can rewrite the evolution of the perturbed field equations (\ref{perturbed-eqm1} and \ref{perturbed-eqm2}) in terms of these new variables. Owing to the suppression of the tachyon energy density in the Starobinsky background, we ignore the perturbation of the tachyon.
Up to the first order in slow-roll parameters, we arrive at,
\begin{align}\label{u-psi}
 u''_\psi+[k^2 -a^2H^2\left(2+3\eta_{\psi}-\epsilon-14\beta^2\frac{1+A^{-2}}{1-A^{-2}}\epsilon\right)]u_\psi &\simeq 0
\end{align}
where we have defined the slow-roll parameters,  $\epsilon=-\frac{\dot{H}}{H^2}$, $\eta=\frac{\dot{\epsilon}}{H\epsilon}$ and $\eta_{\psi}=\frac{w_{,\psi\psi}}{w}$."$'$" denotes the derivative with respect to conformal time $\tau$ which is related to cosmic time by $dt=a\left(t\right)d\tau$
By defining a new variable $y=\frac{k}{aH}$, the equation(\ref{u-psi}) can be re-expressed in terms of $y$ as
\begin{align}\label{u-y}
 y^2\frac{d^2u_\psi}{dy^2} &+\left(1-2p\right) y\frac{du_\psi}{dy}+\{l^2y^2+p^2-\nu^2\}u_\psi=0,
\end{align}
where $l=\left(1-\epsilon\right)^{-1}$, $p=\frac{1}{2}\left(1+\eta\epsilon l^2\right)$, $\nu^2=\frac{3}{2}+\eta_{\psi}+\epsilon-\frac{14}{3}\beta^2\frac{1+A^{-2}}{1-A^{-2}}\epsilon$;
 the second term $p$ is of order two in slow-roll parameters, so we take $p\approx\frac{1}{2}$.
 By solving the above equation and substituting in (\ref{curnature-perturb}), we arrive at co-moving curvature perturbations and the power spectrum. The spectral index is equal to
 \begin{align}\label{ns}
   n_s-1&\simeq 3-2\nu=1-2\epsilon-2\eta_{\psi}-\frac{28}{3}\beta^2\frac{1+A^{-2}}{1-A^{2}} \epsilon.
 \end{align}
The two first terms are the same as usual slow-roll inflation governed by scalaron, but the last term comes from the presence of tachyon. The tensor-to-scalar ratio is, as usual, $r=16\epsilon$.We depicted these observational parameters for various values of model parameters $v_0$ and $\phi_0$ (Fig\ref{Fig5}). These plots show that this model can be compatible with observation for appropriate values of the model parameters.

\section{ Summary and Conclusion}\label{sec-conclude}
We investigate the presence of a tachyon field in a $R+R^2$ background.
By transforming to Einstein frame, the scalaron appears in the $R$ background. Since the Tachyon field is coupled to the scalaron, their evolution has a mutual effect.
To be more precise, we choose a specific potential for the tachyon field (\ref{phi-potential}). During the evolution of fields, the tachyon is suppressed by the scalaron, but the coupling term affects the evolution of the scalaron. The effect of the tachyon can be seen in the spectral index. In (\ref{ns}), the effect of tachyon(the last term) is dominant.

We also check two other common potentials for tachyon field,$V_2\left(\phi\right)=V_0e^{-\left(\frac{\phi}{\phi_0}\right)}$ and $V_3\left(\phi\right)=\frac{V_0}{Cosh\left(\frac{\phi}{\phi_0}\right)}$, it seems the overall behavior is also the same for these potentials.

%~~~~~~~~~~~~~~~~~~~~~~~~~~~~~~~~~~~~~~~~~~~~~~~~~~~~~~~~~~~~~~~~~~~~~~~~~~~~~~~~~~~~~~~~~~~~~~

%~~~~~~~~~~~~~~~~~~~~~~~~~~~~~~~~~~~~~~~~~~~~~~~~~~~~~~~~~~~~~~~~~~~~~~~~~~~~~~~~~~~~~~~~~~~~~~~~~~~

%\section{Acknowledgement}
%\begin{adjustwidth}{-\extralength}{0cm}

%\reftitle{References}

 %%%%%%%%%%%%%%%%%%%%%%%%%%%%%%%%%%%%%%%%%%%%%%%%%%%%%%%%%%%%%%%%%%%

%\bibliography{myref3}

%\PublishersNote{}
%\end{adjustwidth}
\end{document}